# Large magnetoresistance in low temperature metallic region of manganite compounds $(La_{0.7-2x}Eu_x)(Ca_{0.3}Sr_x)MnO_3$ $(0.05 \leq x \leq 0.15)$

D.S. Rana[a], C.M. Thaker[a], K.R. Mavani[a], D.G. Kuberkar[a] and S.K. Malik[b*]

[a]Department of Physics, Saurashtra University, Rajkot - 360 005, India
[b]Tata Institute of Fundamental Research, Mumbai - 400 005, India

**Abstract**

Magnetoresistance (MR) and magnetization (d.c and a.c) measurements have been carried out on the manganites, $(La_{0.7-2x}Eu_x)(Ca_{0.3}Sr_x)MnO_3$ $(0.05 \leq x \leq 0.15)$, in the temperature range of 5K-320K. At 5K, an unusually large MR of almost 98% is observed in the x=0.15 sample, nearly up to fields of 4-5 Tesla. This large high-field MR occurs in the metallic region, far below the insulator-metal transition temperature, and does not vary linearly with applied field. The unusual magnetoresistance is explained the light of various possibilities of phase segregation and cluster spin-glass behavior.
**Keywords:** Manganites, High field MR, Phase segregation.

***Corresponding author**: skm@tifr.res.in (S.K. Malik)

---

The $ABO_3$ type perovskite manganites are being extensively studied due to the possible technological applications of the large magnetoresistance (MR) exhibited by them. A large MR in divalent cation ($Ca^{2+}$, $Sr^{2+}$, $Ba^{2+}$, etc.) doped $RMnO_3$ (R=La, Pr, Nd, etc.) manganites occurs in the vicinity of the insulator-metal (I-M) transition temperature ($T_p$) [1-2]. This magnetoresistance arises from i) intrinsic suppression of magnetic disorder of the Mn-O-Mn couplings and ii) extrinsic effect of the reduction in spin dependent scattering at the grain boundaries [1-4]. The increasing spin polarization below $T_p$ increases the intrinsic magnetic order and conductivity. This causes the MR to fall to very low values as the temperature is lowered below $T_p$. The MR value at low temperatures, far below $T_p$, is almost negligible in single crystals as compared to that in polycrystalline materials suggesting that polycrystalline grain boundaries have a vital role to play in low temperature MR [3]. Hwang *et al.* [4] proposed that, at low temperatures, a nearly complete spin polarized state causes tunneling through grain boundaries and contributes largely to magnetoresistance in low fields. They also showed that there exists a negligible amount of high field MR, varying linearly with field, which is due to spin fluctuations arising as a result of Zener double exchange (ZDE) mechanism.

In this paper, we report magnetoresistance measurements on a new series of manganites, namely, $(La_{0.7-2x}Eu_x)(Ca_{0.3}Sr_x)MnO_3$ (0.05≤x0.15) (to be referred to as LECSMO). This series provides us an opportunity to study the effect of simultaneous increase in size disorder and carrier density by keeping the tolerance factor nearly constant. This is a new and original choice since previously the effect of one factor (keeping others constant) has been studied. Interestingly, in LECSMO series, the tolerance factor 't' is the same as that in the standard [5] $La_{1-x}Ca_xMnO_3$ (x= 0.35, 0.4, 0.45) (LCMO) series, but the other two factors, namely, size-disorder and carrier density, increase with increasing x (t increases marginally from 0.924 for x=0.05 to 0.929 for x=0.15 for LECSMO and from 0.924 for x=0.35 to 0.928 for x=0.45 for LCMO). In LCMO, only the carrier density increases while the size-disorder is negligible in the given range of compositions. We find that in LECSMO samples, in the metallic region and at low temperatures, the high field MR rises significantly, and anomalously reaches nearly a saturated value of 98% in x=0.15 sample. Such a large low temperature MR is an important finding because this is observed in low temperature metallic region where the spin polarization is almost complete.

The $(La_{0.7-2x}Eu_x)(Ca_{0.3}Sr_x)MnO_3$ (0.05≤x0.15) samples were synthesized using the solid-state ceramic preparation route by several grindings and sintering in the temperature range of 1100°C to 1350°C. Powder X-ray diffraction (XRD) measurements were carried out on a Siemens diffractometer using Cu-$K_\alpha$ radiation. DC magnetization measurements were carried out in an applied field of 50 Oe using a SQUID magnetometer (MPMS, Quantum Design). AC susceptibility measurements were performed at different frequencies in an ac field of 5 Oe (PPMS, Quantum Design). Electrical resistivity and magnetoresistance measurements were performed using the four-probe dc technique (PPMS, Quantum Design).

Indexing and Rietveld refinement of XRD patterns of all the $(La_{0.7-2x}Eu_x)(Ca_{0.3}Sr_x)MnO_3$ (0.05≤x0.15) samples revealed that these are single-phase compounds crystallizing in a distorted orthorhombic structure (space group – *Pnma*, No. 62). No lines from impurity phases could be detected within the limits of x-ray detection, which is typically 5%.

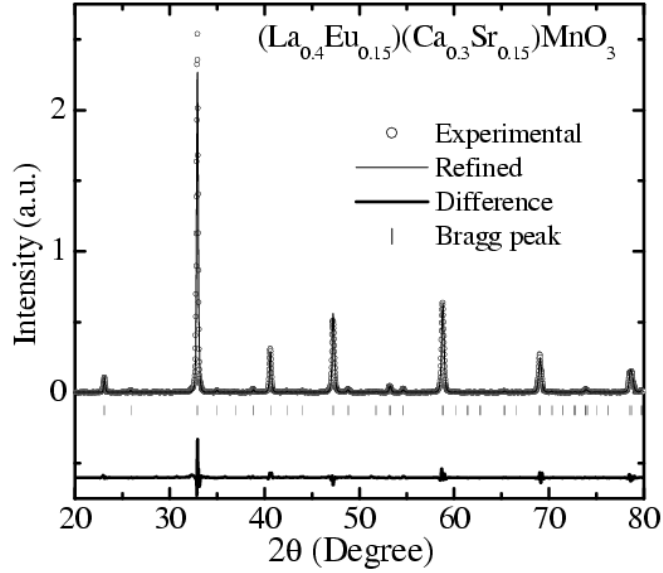

**Fig. 1:** A typical Rietveld fitted XRD pattern for $(La_{0.7-2x}Eu_x)(Ca_{0.3}Sr_x)MnO_3$ (x=0.15) sample.

**Table 1:** Lattice parameters (a,b&c), size-disorder ($\sigma^2$), insulator-metal transition ($T_p$) and Curie temperature ($T_C$) of $(La_{0.7-2x}Eu_x)(Ca_{0.3}Sr_x)MnO_3$ ($0.05 \leq x \leq 0.15$) samples.

| x | a (Å) | b (Å) | c (Å) | $\sigma^2$(Å$^2$) | $T_P$(K) | $T_C$(K) |
|---|---|---|---|---|---|---|
| 0.05 | 5.547(1) | 7.708(1) | 5.465(3) | 0.0012 | 175 | 195 |
| 0.10 | 5.456(4) | 7.702(3) | 5.458(2) | 0.0021 | 158 | 175 |
| 0.15 | 5.442(4) | 7.681(2) | 5.447(1) | 0.0030 | 78 | 130 |

Figure 1 displays a typical Rietveld fitted XRD pattern for x=0.15 sample. The lattice parameters, obtained from the above-mentioned refinement, are given in Table 1 from which it is seen that the cell volume decreases with increasing Eu content, x. In this series of compounds, with increasing substitution of $Eu^{3+}$ (atomic radius 1.12Å) and $Sr^{2+}$ (1.31Å), the average A-site cation radius, $<r_A>$, remains nearly constant throughout the series, while the size-disorder and the carrier density increase. Therefore, the decrease in cell volume may be attributed to increasing $Mn^{4+}$ concentration (due to increasing $Sr^{2+}$ concentration) and the lattice distortion caused by increasing size disorder (size disorder is given by $\sigma^2 = \Sigma x_i r_i^2 - <r_A>^2$) [6].

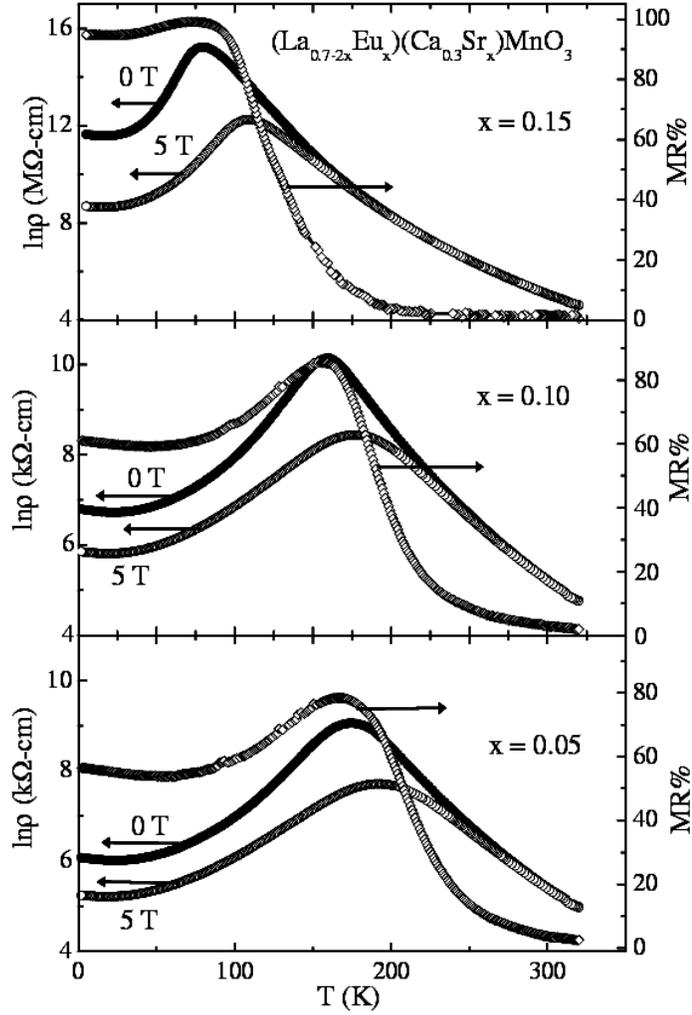

**Fig. 2:** Resistivity (ρ) vs. temperature (T) plots in 0T and 5T fields and MR% vs. temperature (T) in 5T plot for $(La_{0.7-2x}Eu_x)(Ca_{0.3}Sr_x)MnO_3$ (0.05≤x0.15) samples.

Figure 2 shows the electrical resistivity (ρ) versus temperature (T) plots for $(La_{0.7-2x}Eu_x)(Ca_{0.3}Sr_x)MnO_3$ (0.05≤x0.15) samples in zero applied magnetic field as well as in a field of 5T. This figure also shows magnetoresistance (magnetoresistance is defined as MR%=$(\rho_0-\rho_H)/\rho_0)\times100$) as a function of temperature in a field of 5T. The I-M transition temperature, $T_p$, falls from 175K for x=0.05 sample to 79K for x=0.15 sample. The fall in $T_p$ may be attributed to the increased Coloumbic interactions (due to increased carrier density) and to the increased size disorder at the A-site. It may be seen that the MR is maximum in the vicinity of $T_p$ for x=0.05 and x=0.10 samples, whereas it is almost saturated from $T_p$ down to 5K for x=0.15 sample. The MR of nearly 98% at 5K for x=0.15 sample is unusual and hence interesting. At low temperatures, the high field MR observed in the present samples is even larger than the earlier reported low temperature MR of ~ 60-70% in other polycrystalline manganite systems (7-9).

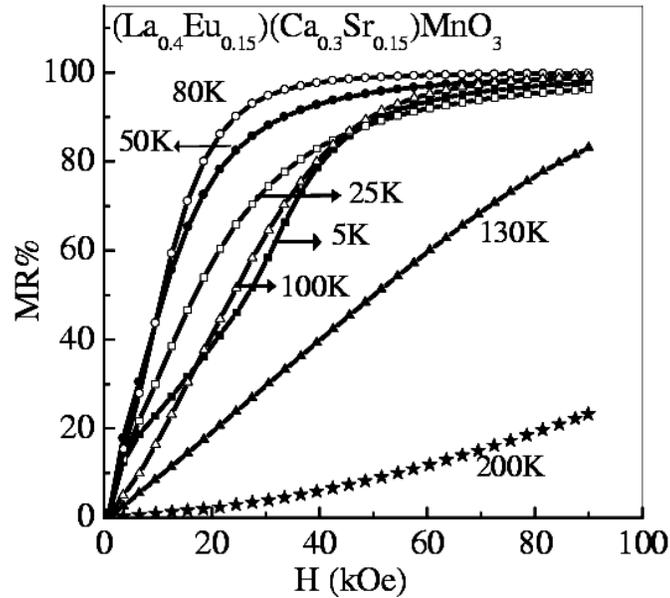

**Fig. 3:** MR% vs. applied magnetic field (H) plots for $(La_{0.4}Eu_{0.15})(Ca_{0.3}Sr_{0.15})MnO_3$ sample.

To explore the MR behavior of x=0.15 sample in detail, we have studied the variation of MR with applied magnetic field in this sample and the results are shown in Fig. 3. As expected, the MR is largest in the vicinity of $T_p$ for all the samples (data not shown for x=0.05 and 0.10 samples for the sake of brevity). The MR isotherms at temperature lower than $T_p$ also display large MR and these isotherms behave differently than those in the vicinity of the peak in resistivity. Here it may be mentioned that, at low temperatures, a low field MR occurs due to inter-grain spin polarized tunneling of the carriers, whereas the high field MR has its origin in ZDE and reduction in grain boundary scattering.

We have plotted the MR isotherms for all the $(La_{0.7-2x}Eu_x)(Ca_{0.3}Sr_x)MnO_3$ (0.05≤x0.15) samples at 5K, separately, in Fig. 4. The MR behavior in these compounds may be divided into two parts. The MR of nearly 20% below a field of ~0.5T may be termed as the low field MR while the MR from 20% to 98% in fields of 0.5T to 9T may be called the high field MR. The low field MR is identified as the linear part of the isotherms in low fields (< 1Tesla) and is due to spin polarized tunneling while the rest of the contribution in MR isotherms is due to high field [4]. It is clear that, in the presently studied samples, the high field MR is appreciably larger than the low field MR. In particular, for x=0.15 sample, the high field MR is almost five times larger than the low field MR. Also, the high field MR in this sample shows a large deviation from the linear relationship of MR with applied field.

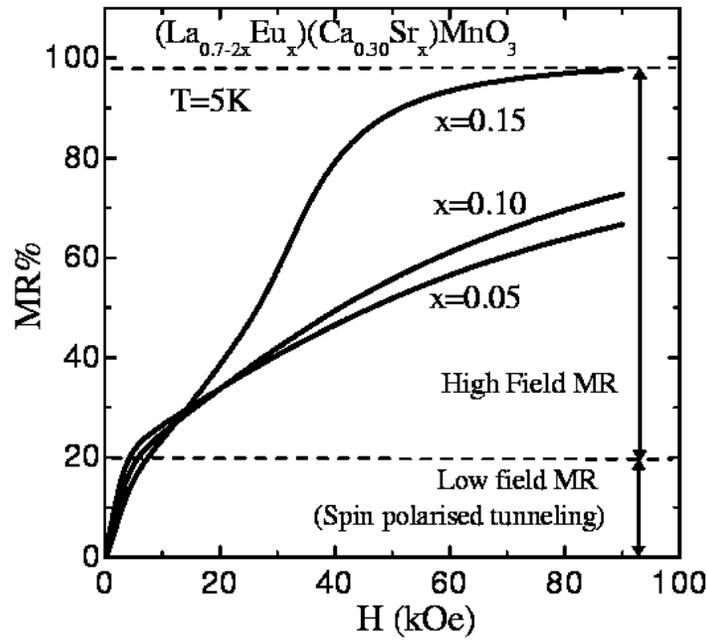

**Fig. 4:** MR% vs. applied magnetic field (H) plots for $(La_{0.7-2x}Eu_x)(Ca_{0.3}Sr_x)MnO_3$ $(0.05 \leq x \leq 0.15)$ samples at 5K.

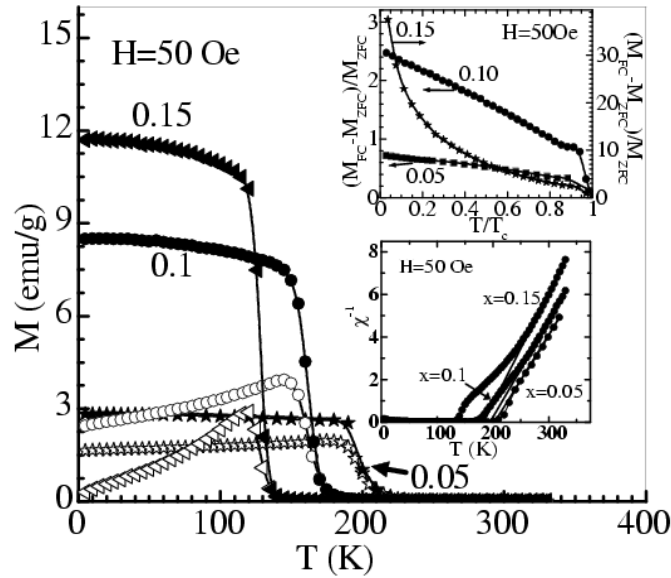

**Fig. 5:** Zero-field-cooled (ZFC-hollow symbols) and field-cooled (FC-solid symbols) magnetization (M) vs. temperature plots (T) for $(La_{0.7-2x}Eu_x)(Ca_{0.3}Sr_x)MnO_3$ $(0.05 \leq x \leq 0.15)$ samples. Inset figure shows the fractional change in magnetization vs. normalized temperature $(T/T_C)$ and inverse susceptibility $(\chi^{-1})$ vs. temperature plots for same samples.

Figure 5 shows the zero-field cooled (ZFC) and field cooled (FC) magnetization (M) versus temperature (T) plots for all the $(La_{0.7-2x}Eu_x)(Ca_{0.3}Sr_x)MnO_3$ $(0.05 \leq x \leq 0.15)$ samples. The $T_C$ (determined from the bifurcation of the $M_{ZFC}$ and $M_{FC}$ curves) decreases from 195K for x=0.05 to 130K for x=0.15 sample. Insets in the figure show the fractional changes in magnetization (calculated as {$M_{FC}$-$M_{ZFC}$}/$M_{ZFC}$) versus T/$T_C$ and inverse susceptibility versus temperature plots for all the $(La_{0.7-2x}Eu_x)(Ca_{0.3}Sr_x)MnO_3$ (0.05≤x≤0.15) samples. The {$M_{FC}$-$M_{ZFC}$}/$M_{ZFC}$ vs. T/$T_C$ plot is significant to emphasize the extent of separation of $M_{ZFC}$ and $M_{FC}$ curves. With decreasing temperature, this separation increases rapidly for x=0.15 sample and points towards the cluster-spin glass behavior of this sample. This has been verified by measuring the frequency dependence of the a.c. susceptibility.

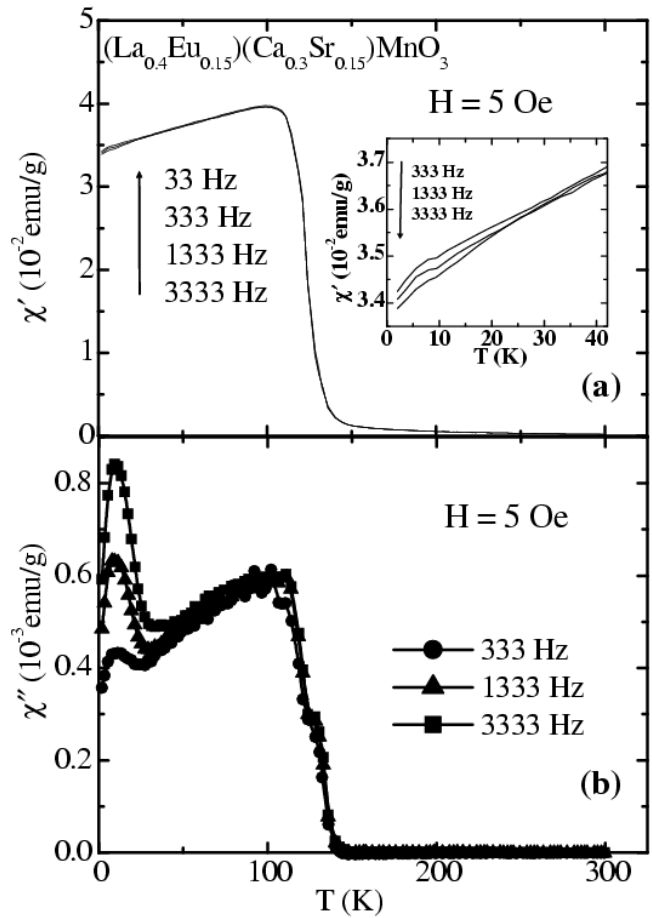

**Fig. 6:** (a) Real a.c. susceptibility ($\chi'$) and (b) imaginary a.c. susceptibility ($\chi''$) vs. temperature (T) at different frequencies for $(La_{0.4}Eu_{0.15})(Ca_{0.3}Sr_{0.15})MnO_3$ sample. Inset figure (a) shows the magnified part of low temperature real a.c. susceptibility ($\chi'$) for the same sample.

Figure 6 [(a) and (b)] shows the real and imaginary parts of a.c. susceptibility ($\chi'$ & $\chi''$, respectively, in an ac field of 5 Oe) vs. temperature at frequencies of 333, 1333 and 3333 Hz for the x=0.15 sample. There is a weak frequency dependence of $\chi'$ at T<30 K whereas, at T>30 K,

the $\chi'$ behaves in a similar way at all frequencies. Interestingly, this weak frequency dependence of $\chi'$ is accompanied by a large frequency dependence of $\chi''$ below 30 K. It is seen that two peaks appear in $\chi''$ vs. T plots. The first peak at about ~130 K corresponds to the paramagnetic to ferromagnetic transition (with no frequency dependence). The second peak in $\chi''$ has pronounced frequency dependence and shifts from 9 K at 333 Hz to 13 K at 3333 Hz. This is indicative of the cluster spin-glass like behavior. This type of cluster spin-glass behavior, evidenced from a second peak below $T_C$ in $\chi''$, has also been reported for cobaltites [10] and manganites [11].

The dc magnetization (M) for x=0.15 sample has been measured as a function of magnetic field (H) at 5K (plot not shown) in which a saturation in magnetization (>3$\mu_B$) is obtained in a field of 5 Tesla. This experimental saturation magnetization value is close to that calculated from the spin only value (3.55$\mu_B$) of Mn ions, which implies that there is no disorder of Mn spins. Therefore, the origin of the large MR cannot be attributed to the intrinsic factor of Zener-Double Exchange. However, it is seen from Fig. 4 that, for the x=0.15 sample, there is large deviation in paramagnetic susceptibility from the Curie-Weiss behavior. This points towards the segregation of both ferromagnetic-metallic and paramagnetic-insulating phases from the major matrix. Our assumption of phase segregation is also supported by a large disparity between $T_C$ (~130K) and $T_p$ (~80K) and the cluster spin-glass behavior for x=0.15 sample. Phase segregation, as a result of disparity between $T_p$ and $T_C$, has also been reported to exist in similar compounds having large size disorder [12,13]. This phase segregation may have a strong correlation with the high field MR at low temperatures.

It has also been reported that connectivity between grains, grain boundary (GB) contaminations and pinning of Mn spins at GB play an important role in the conduction of carriers [7-9]. The high field MR is due to the opening up of new conduction channels on GBs. These conducting channels open linearly with applied field and, therefore, at higher fields the MR varies almost linearly with field. In the present case, the unusually large MR observed in x=0.15 sample, at 5K, in higher fields may also be explained on the basis of GB contamination, pinning of Mn ion spins at GB and by the possible existence of microscopic insulating phases embedded between connecting grains. These factors create anisotropic environment of insulating and conducting channels around Mn ions and blocks the path for carriers at the interfaces of connecting grains. The large size disorder and phase segregation in the present samples largely increases the weight of the insulating channels. However, in large applied magnetic field, such insulating channels may start conducting and may allow the carriers to percolate by opening of new conduction paths. But, the opening up of new conduction paths is not linear with the applied field. Also, we have evidenced a weak cluster spin-glass behavior (fig. 6) in the temperature range where a large anomalous MR% occurs. This cluster spin-glass behavior may be a result of phase-segregation in x=0.15 sample. The frozen cluster magnetic moments may align with large applied magnetic fields leading to large high field MR at low temperatures.

In summary, magnetoresistance behavior of $(La_{0.7-2x}Eu_x)(Ca_{0.3}Sr_x)MnO_3$ (0.05≤x0.15) samples has been studied. Due to large size-disorder in these compounds, the $T_C$ and $T_p$ fall more rapidly than that of standard LCMO [5] compounds. At low temperatures, in the metallic region, an unusually large high field MR ~ 98% is observed in x=0.15 sample. Size-disorder induced features such as cluster-glass behavior; possible phase segregation and pinning of Mn ion spins are the factors responsible for such a low temperature MR. Application of high fields result in the opening of conduction paths in a non-linear way for blocked Mn spins at the polycrystalline grain boundaries and the alignment of frozen cluster magnetic moments. This study opens up a new way to enhance the low temperature magnetoresistance by means of inducing large size-disorder in heavily hole-doped manganites.

## ACKNOWLEDGEMENT

DGK is thankful to University Grants Commission (UGC), New Delhi, India for financial support in the form of a major research project. DSR is thankful to CSIR, India for providing Senior Research Fellowship (SRF).

**References:**

1. "Colossal magnetoresistance, charge ordering and other related properties of rare earth manganates" edited by C.N.R. Rao and B. Raveau, World Scientific Publishing Company (1998).
2. R. Von Helmolt, J. Wecker, B. Holzapfel, L. Schultz, K. Samwer, Phys. Rev. Lett. 71, 2331 (1993).
3. A. Gupta, G. Q. Gong, G. Xiao, P. R. Duncombe, P. Lecoeur, P. Trouilloud, Y. Y. Wang, V. P. Dravid and J. Z. Sun, Phys. Rev. B 54, 15629 (1996).
4. H. Y. Hwang, S.-W Cheog, N. P. Ong, and B. Batlogg, Phys. Rev. Lett. 77, 2041 (1996).
5. P. Schiffer, A. P. Ramirez, W. Bao, and S-W. Cheong, Phys. Rev. Lett. 75, 3336 (1995).
6. L.M. Rodriguez-Martinez, J.P. Attfield, Phys. Rev. B 54, 15622 (1996).
   L.M. Rodriguez-Martinez, J.P. Attfield, Phys Rev. B 58, 2426 (1998).
7. A. de Andres, M. Garcia-Hernandez, J. L. Martinez, C. Preito, Appl. Phys. Letts. 74, 3884 (1999) and references therein.
8. A. de Andres, M. Garcia-Hernandez, J. L. Martinez, Phys. Rev. B 60, 7328 (1999) and references therein.
9. Sheng Ju, H. Sun and Zhen-Ya Li, Phys. Lett. A 300, 666 (2002).
10. D.N.H. Nam, K. Jonason, P. Nordblad, N.V. Kheim, N.X. Phuc, Phys. Rev. B 59, 4189 (1999).
11. L. Ghivelder, I.A. Castillo, M.A. Alonso, L.F. Cohen, Phys. Rev. B 60, 12184 (1999)
12. Hirotoshi Terashita and J.J. Neumeier, Phys. Rev. B 63, 174436 (2001).
13. Young Sun, M.B. Salamon, Wei Tong, Yuheng Zhang, Phys. Rev. B 66, 94414 (2002).